\documentclass
[aps,twocolumn,floats,balancelastpage,showpacs,preprintnumbers,prd,amssymb,floatfix,nofootinbib,superscriptaddress,amsmath]{revtex4}

\usepackage{epsfig}
\usepackage{epstopdf}

\usepackage{float}
\usepackage[usenames]{color}
\usepackage{color,epsfig,amsmath,amssymb,fancyhdr}   
\usepackage[french]{babel}
\usepackage{ulem}   

\newcommand{\beq}{\begin{equation}}
\newcommand{\eeq}{\end{equation}}
\newcommand{\ben}{\begin{eqnarray}}
\newcommand{\een}{\end{eqnarray}}
\newcommand{\bi}{\begin{itemize}}
\newcommand{\ei}{\end{itemize}}

\begin{document}
\preprint{IRFU-10-185}

\title{Searches for dark matter subhaloes with wide-field Cherenkov telescope surveys}

\author{Pierre Brun}
\email{pierre.brun@cea.fr}
\affiliation{CEA, Irfu, Service de Physique des Particules, Centre de Saclay, F-91191
Gif-sur-Yvette | France}

\author{Emmanuel Moulin}
\email{emmanuel.moulin@cea.fr}
\affiliation{CEA, Irfu, Service de Physique des Particules, Centre de Saclay, F-91191
Gif-sur-Yvette | France}

\author{J\"urg Diemand}
\email{diemand@physik.uzh.ch}
\affiliation{Institute for Theoretical Physics, University of ZŸurich, CH-8057
ZuŸrich | Switzerland}

\author{Jean-Fran\c cois Glicenstein}
\email{jean-francois.glicenstein@cea.fr}
\affiliation{CEA, Irfu, Service de Physique des Particules, Centre de Saclay, F-91191
Gif-sur-Yvette | France}

\begin{abstract}

The presence of substructures in dark matter haloes is an unavoidable consequence of the cold dark matter paradigm. Indirect signals from these objects have been extensively searched for  with cosmic rays and $\gamma$ rays. At first sight, Cherenkov telescopes seem not very well suited for such  searches, due to their small fields of view and the random nature of the possible dark matter substructure positions in the sky.  However, with long enough exposure and an adequate observation strategy, the very good sensitivity of  this experimental technique allows us to constrain particle dark matter models. We confront here the sensitivity map of the HESS experiment built out of their Galactic scan survey to the state-of-the-art cosmological N-body simulation Via Lactea II. We obtain competitive constraints on the annihilation cross section, at the level of 10$^{-24}$$-$10$^{-23}$ cm$^{3}$ s$^{-1}$. The results are extrapolated to the future Cherenkov Telescope Array, in the cases of a Galactic plane survey and of an even wider extragalactic survey.  In the latter case, it is shown that the sensitivity of the Cherenkov Telescope Array will be sufficient to reach the most natural particle dark matter models.
\end{abstract}


\maketitle

\section{Introduction}

In the current cosmological paradigm, cold dark matter (CDM) dominates structure formation. The haloes of galaxies and clusters of galaxies are assembled through the merging of a huge number of smaller structures. Most mergers are incomplete and large CDM haloes, {\it e.g.} the one around the Milky Way, harbor an enormous population of subhaloes, which  are a record of its assembly history. Some particle physics models beyond the standard model predict the existence of these new weakly interacting massive particles  (WIMPs) that could well form the cosmological dark matter (DM).  Should these particles be thermally produced in the early Universe, their comoving density would have been regulated by self-annihilations. The freeze-out of this reaction in the early Universe leads to the establishment of the nowadays observed CDM density. In dense enough structures, the very same annihilation process can efficiently convert particle DM mass energy into high energy standard model particles.  The CDM structures represent at least one-tenth of the total halo mass and are  privileged targets of searches for DM particle annihilations. The dense regions worth considering are the Galactic center and galactic subhaloes, including a small subset of them which harbors the dwarf satellite galaxies of the Milky Way. The principle of indirect searches for DM through $\gamma$ rays is to search for the $\gamma$-ray emission following the hadronization and/or decays of these exotically produced standard model particles.

Strategies for searching this dim radiation include targeted searches or wide-field surveys. Thanks to their very large collection area, imaging atmospheric Cherenkov telescopes (IACT) are very well suited for deep observations of selected sources. Constraints on particle DM models have been obtained by the HESS experiment from the observation of the Galactic center~\cite{Aharonian:2006wh}, the Sagittarius dwarf galaxy~\cite{Aharonian:2007km}, and the Canis Major overdensity~\cite{2009ApJ...691..175A} and by MAGIC from the observation of the Perseus galaxy cluster~\cite{Aleksic:2009ir} and Milky Way satellites Draco~\cite{Albert:2007xg} and Willman 1~\cite{Aliu:2008ny}. The VERITAS experiment obtained constraints from the observation of Draco, Ursa Minor, Bo\"{o}tes 1, and Willman 1~\cite{veritas:2010pja} and the Whipple Collaboration got constraints from Draco, Ursa Minor and M15~\cite{Wood:2008hx}. In this paper we focus instead on nontargeted searches with Cherenkov telescopes. Space-based instruments such as the Fermi satellite can much more easily perform blind searches for DM subhaloes with a regular scanning of the entire sky thanks to their large field of view. Some prospect regarding searches with Fermi have been conducted in Refs.~\cite{Anderson:2010df, Pieri:2009je, Giocoli:2007gf} and recently the first analyses based on actual data were released (Refs.~\cite{Buckley:2010vg, Mirabal:2010ny}).
In~\cite{Aharonian:2008wt}, HESS data from the Galactic plane survey have been used to perform for the first time a blind search for DM substructures with a wide-field survey with IACTs (the structures were DM spikes around intermediate-mass black holes ). However, this substructure scenario is rather optimistic since the abundance and the properties of intermediate-mass black holes and of the DM spikes around them remain practically unconstrained. Here we investigate constraints derived using the HESS Galactic survey and the conventional CDM subhalo distribution obtained by the cosmological N-body simulation Via Lactea II (VL-II)~\cite{vl2}. As we shall see in the following, the outcome of this study is based only on the numerically well resolved distribution of CDM structures in the Galactic halo, it is thus quite robust. Also, it does not rely on further density enhancements such as $e.g.$ the possible formation of intermediate-mass black holes. This results in weaker but safer constraints on the particle physics models than those of ~\cite{Aharonian:2008wt}. To go beyond  these constraints, it is interesting to consider extended surveys that will certainly be performed by the next generation of IACT such as the Cherenkov Telescope Array (CTA)~\cite{cta} as high priority observations. We thus extend the study  by extrapolating the current constraints to the sensitivity of  CTA, the future observatory in this range of energy.

The paper is structured as follows. In Sec.~\ref{SubhaloContent} we present the results of  the VL-II simulation regarding the subhalo population which we are interested in. Section~\ref{SensitivityMaps} is devoted to the description of the HESS sensitivity map and how the extrapolation to the CTA is done. The results inferred from the HESS survey are presented in Sec.~\ref{Results}. Finally, Sec.~\ref{prospects} presents the prospects with the CTA and Sec.~\ref{Conclusions} is devoted to the conclusion.

\section{Predictions for the subhalo content of the Milky Way}
\label{SubhaloContent}

The CDM subhalo distribution is taken directly from the VL-II simulation \cite{vl2}, one of the largest, most accurate cosmological N-body simulations of the Galactic CDM halo. The particle mass of 4100 $M_{\odot}$ allows us to resolve small subhaloes ($> 10^5 M_{\odot}$) throughout the Galactic halo and even as close to the Galactic center as the solar neighborhood. Much smaller CDM subhaloes are expected to survive as well~\cite{Diemand:2005vz}, but we will show now that for the purposes of this analysis it is not necessary to include CDM clumps below the VL-II resolution limit.  The subhaloes resolved in VL-II increase the luminosity of the entire haloes as seen by a distant observer by a factor of 1.9. Including the smaller unresolved haloes expected to exist in CDM does increase this factor to values between 4 and 15, depending on the abundance and properties assumed for these smaller clumps. The total luminosity from subhaloes is also increased by similar factors due to their own subsubstructrue~\cite{Anderson:2010df}, but we assume smooth subhaloes throughout this work for simplicity. Locally (within 1 kpc of the Solar System) this substructure models leads to an enhancement of only 1.4$\pm$0.2~\cite{vl2}.

The dark matter annihilation luminosity\footnote{Here we assume a smooth subhalo. Clumps within a
subhalo do enhance the luminosity~\cite{Anderson:2010df}, but are
ignored here.} is defined by
\beq
L\;=\;\int_{\rm V} \frac{\rho^2}{2} dV\;\;,
\eeq
where $\rho^2$ is the squared mass density of the subhalo.
The smallest clumps, which are still well enough resolved in VL-II, have peak circular velocities of $V_{\rm max} = 3 \; \rm km\, s^{-1}$ and 
$L=1.7 \times 10^5 \rm  M_{\odot}^2 pc^{-3}$
, and their local  separation is $\langle d \rangle = 5.8$ kpc; {\it i.e.} in a random realization they are found at a median distance of $D_0 \simeq 0.5 \langle d \rangle \simeq 2.9$ kpc from the observer~\cite{Brun:2009aj}. The $\gamma$-ray flux from a given clump is calculated from
\beq
\Phi(>E_{\rm th})\;=\;\frac{1}{4\pi D^2}\;\;\frac{\sigma v}{m^2}\;\;L\;\;N_{\gamma}(>E_{\rm th})\;\;,
\eeq
$N_{\gamma}$ being the integrated number of $\gamma$-rays produced in a WIMP collision above a given energy threshold $E_{\rm th}$, $m$ the WIMP mass, and 
$\sigma v$ the velocity-weighted annihilation cross section. With $D = 2.9$ kpc, one gets the reference flux of $\Phi_0=\rm 7\times10^{-14} \; cm^{-2} s^{-1}$ from such a small, but still well resolved, VL-II clump .
This value is obtained for $\sigma v= 3\times 10^{-26} \rm~cm^3 s^{-1}$, $m$ = 500~GeV in the case of an annihilation into $\tau$ pairs.
As we shall see in the following, the observable flux for HESS is roughly $\Phi_{\rm HESS}\;=\;10^{-12}\;{\rm cm^{-2}s^{-1}}\;=\;  14 \times \Phi_0$ and in the best case for the CTA is $\sim 3\times 10^{-13}\;{\rm cm^{-2}s^{-1}}$, which is still larger than $\Phi_0$. This means that the smallest clumps that the VL-II simulation is able to resolve are already too faint to be observed by HESS or the CTA for most random realizations. They would need to fall unusually close to the observer to be detected ($D < 1.8$ kpc), which happens in only 18\% of all realizations. A HESS detection of a smaller CDM clump, below the numerical resolution of our simulation is even less likely: In CDM the distance $D_{\rm n}$ to the nearest subhalo of above given luminosity scales roughly like $D_{\rm n}\;\propto\;L^{\rm 1/3}$(see for instance~\cite{Brun:2009aj}). The flux $\Phi \propto L/D^2$ therefore goes like $\Phi\ \propto L^{1/3}$, {\it i.e.} the flux from the nearest subhalo in a given bin in log$L$ (or log subhalo mass) increases with subhalo mass and we can safely ignore small subhaloes below the VL-II numerical resolution for  the present analysis.

\section{Flux sensitivity maps}
\label{SensitivityMaps}

\begin{figure*}[t]
\centering
\includegraphics[width=2.1\columnwidth]{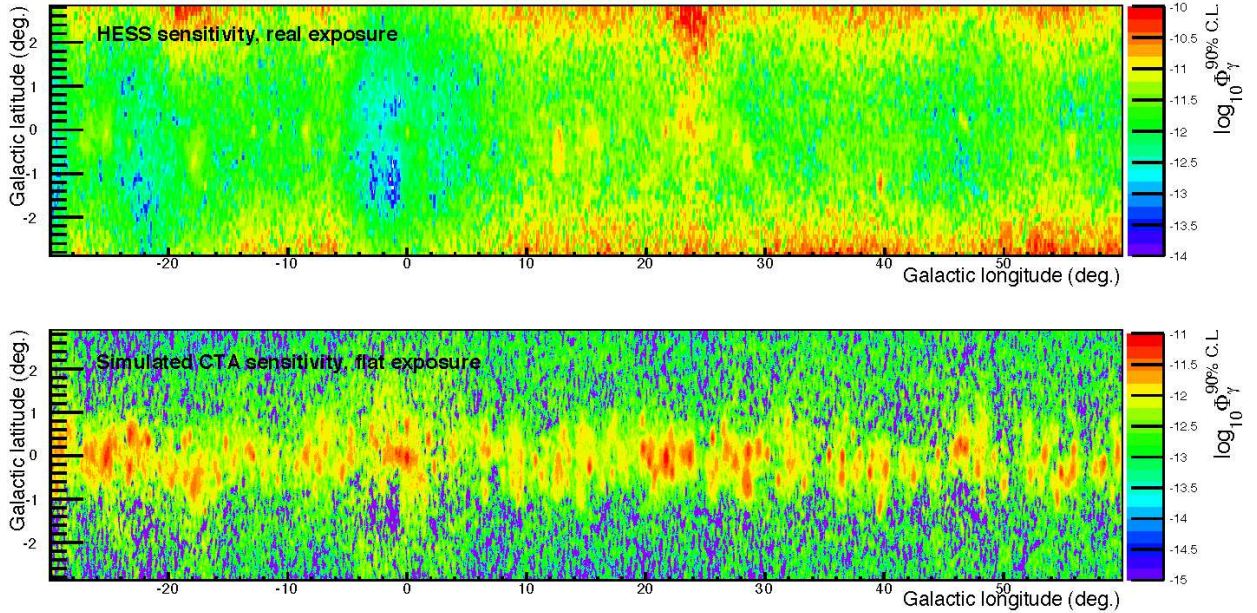}
\caption{Flux sensitivity maps at 90\% C.L. for HESS  and a CTA-like array. The maps are calculated here for a 500~GeV DM particle annihilating 
with 100\% BR into $b\bar{b}$. Top: The  HESS  sensitivity map is calculated with real exposure. Bottom: The simulated flux sensitivity map for a CTA-like array is obtained for a flat exposure of 10h. \label{maps}}
\end{figure*}

Data from the Galactic plane survey conducted by the HESS experiment allowed us to obtain flux sensitivity maps to DM annihilations~\cite{Aharonian:2008wt}. 
In Ref.~\cite{Aharonian:2008wt}, the $\gamma$-ray flux sensitivity is calculated in each position depending on the acceptance of the detector, the exposure to each point of this region of the sky and the predicted $\gamma$-ray spectrum, according to
 \begin{eqnarray}
\label{eqn:phi} \lefteqn{\Phi_{\gamma}^{90\% C.L}(b,l)\,=\,}
\nonumber
& & \\
& &
\hspace{-1.0cm}\frac{N_{\gamma}^{90\%C.L}(b,l)\,\displaystyle N(>E_{\rm th})}
{\displaystyle{T_{\rm obs}(b,l)}\int_{0}^{m} A_{\rm
eff}(E_{\gamma},b,l)\frac{dN}{dE_{\gamma}}(E_{\gamma})\,dE_{\gamma}}.
\end{eqnarray}
Here, A$_{\rm eff}$ is the effective area for $\gamma$ rays, which is a function of the gamma energy E$_{\gamma}$, the sky position $(b,l)$, and T$_{\rm obs}$
the observation time at a given sky position. Notice that the flux sensitivity depends explicitly  on the mass of the DM particle $m$ and on the details of the annihilation process. 
The HESS Galactic plane survey allowed us to detect a large population of unidentified sources~\cite{Aharonian:2005jn,2006ApJ...636..777A}.
None of them exhibit an energy cut-off in the covered energy range. Each $\gamma$-ray source is best-characterized by a power-law spectrum, as expected for $\gamma$-ray emission from standard acceleration astrophysical processes.  Consequently it excludes them from being relevant candidates for DM clumps.
No further information about the energy spectrum is used because the nondetection of DM clumps --which is the baseline assumption throughout this paper-- depends first on the total flux. As far as the integrated flux is used, the energy spectrum of the searched sources only influences the results indirectly though the choice of the annihilation channel.
Fig.~\ref{maps} shows the HESS flux sensitivity map to DM annihilation for a 500 GeV DM particle annihilating into $b\bar{b}$. In the following,
when other annihilation channels or masses are considered, the flux limit is properly rescaled. The sensitivity map contains hints of the presence of all the discovered sources by HESS. As expected, the flux sensitivity decreases at the position of the detected sources. The flux sensitivity map can be easily understood as follows: if a $\gamma$-ray source at a given Galactic position gave a larger flux than the value quoted in the corresponding bin, it would
have been detected. Consequently, any model predicting a statistically significant number of sources of any kind with fluxes above these values is excluded.

An important point concerns diffuse emission, from either cosmic rays, the smooth DM halo or unresolved clumps. Note that the Cherenkov technique is suitable for detecting pointlike sources and slightly extended sources or diffuse emission. The reason is that one has to perform an on-off type of background subtraction (see for instance ~\cite{Aharonian:2006pe}). Because the trigger rate can vary from one observation to another (depending on the observing conditions, the zenith angle, night sky background level, etc.), the angular scale one has to keep in mind is the size of the effective field of view of an individual run, which is about $4^{\circ}\times 4^{\circ}$\footnote{On the CTA, the individual field of view will be larger ($>7^\circ$) for energies above 100 GeV, but this does not qualitatively affect the discussion on diffuse emission}. Unless a more subtle method is used,  the off region has to be taken in the same field of view as the on region to avoid large systematic errors. This means that if a signal is spread over a region that is larger than $5^\circ$, it cannot be observed by Cherenkov telescopes using the on-off technique. Although they concern large fields of view, the maps of Fig.~\ref{maps} are built out of a concatenation of smaller $4^{\circ}\times 4^{\circ}$ observations. Concerning diffuse emissions on scales below $\sim 4^{\circ}$, some has been detected by HESS ($e.g.$ in the Galactic center region~\cite{Aharonian:2006au}) and is an additional source of background, the point-source sensitivity within those regions is decreased. This is clearly visible when scrutinizing the map of Fig.~\ref{maps}. Any diffuse signal that would be pretty much constant all over the map --should it reach or even exceed the point-source sensitivity of the experiment-- would actually be erased by the background subtraction procedure.
 In practice it is always the case; for instance the signal from the smooth halo and unresolved clumps vary by a factor of at most $\sim 20\%$ on a $4^{\circ}$ scale~\cite{Pieri:2007ir}. The only region where it can be untrue is the central $1^\circ$ where the central cusp of the DM halo could appear as a slightly extended source, hence producing a large gradient diffuse emission. This is not a concern for our study, because of the presence of the central astrophysical source. The issue of an annihilation signal at the Galactic center has been addressed anyway in~\cite{Aharonian:2006wh}. In a sense, because of the way the sensitivity map is built, the issue of large scale diffuse emission is handled automatically. This is true as long as the sources that are searched for have an extension lower than $\sim2^{\circ}$ . As we shall see in the following, it is always so for the considered case.

The next generation of IACT will consist of a large telescope array (the CTA)~\cite{Consortium:2010bc}.  The  current effort  on its design should allow us to improve significantly the global performance of the present generation. The goal is to extend the accessible energy range towards both the low and the high energies and gain at least a factor of 10 in flux sensitivity. To extrapolate the current results to future observatories, different assumptions are made regarding the foreseen CTA characteristics. A conservative value of  50 GeV for the energy threshold is assumed. The effective area of the instrument is improved by a factor of 10 and the hadron rejection by a factor  of  2.  The overall improved capabilities of the CTA will presumably allow us to detect a number of new $\gamma$-ray sources. The  construction of the CTA-extrapolated map follows the procedure described in~\cite{Funk:2009kp}. Assuming a supernovae remnant source model for the $\gamma$-ray emission~\cite{Drury:1993pd} and a radial source distribution~\cite{1996A&AS..120C.437C}, the distance and the $\gamma$-ray flux are calculated. The Galactic plane is then randomly populated according to  the spatial distribution of sources observed by the HESS survey~\cite{Aharonian:2005jn,2006ApJ...636..777A}. So the HESS $\gamma$-ray flux source distribution is extrapolated to CTA performance. This extrapolation results in the prediction for the discovery of a few hundreds of new sources in the Galactic survey field of view. The presence of  these 
new sources deteriorates the DM flux sensitivity accordingly. The  resulting projected CTA sensitivity map is shown in the lower panel of Fig.~\ref{maps} for a DM particle mass of 500 GeV annihilating into $b\bar{b}$. A flat exposure of 10 h in each position of the map is assumed. This value allows us to match the total amount of time for the  CTA survey to the   $\sim$ 400 h which were needed by HESS to survey this region of the sky. The flux sensitivity for CTA ranges from $\sim$$\rm 10^{-12}\,cm^{-2}s^{-1}$ in the region where the new sources are present to a few $\rm 10^{-13}\,cm^{-2}s^{-1}$ on average at higher latitudes.

\section{Current exclusion limits}
\label{Results}

\begin{figure}[t]
\centering
\includegraphics[width=.8\columnwidth]{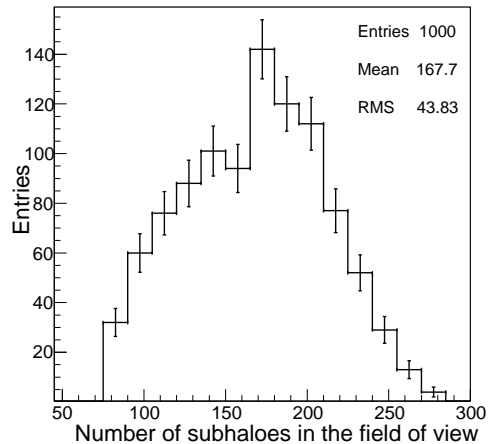}
\caption{Distribution of the number of subhaloes  between $\pm$3$^{\circ}$ in Galactic latitude and from -30$^{\circ}$ to 60$^{\circ}$ in
Galactic longitude.  The distribution is obtained from 1000 stochastic realizations of the Milky Way-like halo from VL-II. The mean of the distribution is 168 with a rms  of 44.
\label{nbclumps}}
\end{figure}

The question  to be addressed in this section is --for a fixed set of particle physics parameters-- what the probability is for a clump to lie in the survey region with a flux larger than the HESS sensitivity at its position in the sky. To answer this question, we generate $10^3$ different Monte Carlo realizations of the Milky Way halo. Independent realizations are obtained by randomly placing the observer at a distance of 8.5 kpc from the Galactic center. Over our realizations, the total number of subhaloes inside the Galactic survey field of view shown in Fig.~\ref{maps} is $168\pm44$, out of the $\sim 10^4$ resolved subhaloes contained in the Milky Way. The distribution obtained after these $10^3$ virtual experiments is displayed in Fig.~\ref{nbclumps}. The probability to find no subhalo in the field of view is  less than 10$^{-3}$. The spatial distribution of clumps is slightly triaxial, with an unknown orientation which is not expected to be correlated with the baryonic distribution. For that reason, the distribution is wider than what is expected from a purely spherical distribution. To scan the  (DM particle mass--$\sigma v$) plane, the mass of the DM particle is  kept fixed and the value of the cross section for which 2.3  subhaloes are visible {\it on average} is searched. This corresponds to a 90\% C.L. limit on $\sigma v$.

\begin{figure}[t]
\centering
\includegraphics[width=.8\columnwidth]{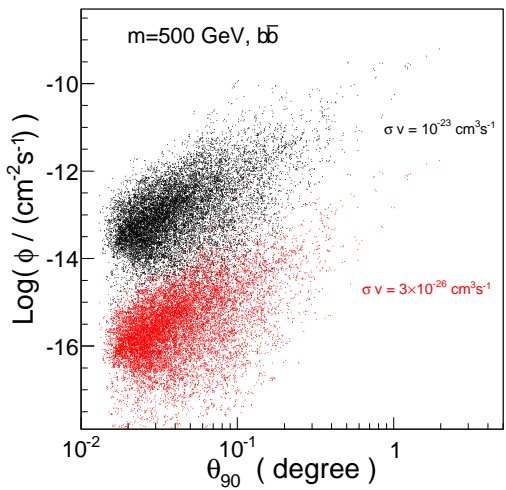}
\caption{ Distributions of flux and angular sizes of subhaloes in the HESS field of view, obtained for annihilation cross sections of $10^{-23}$ and $3\times 10^{-26}\;\rm cm^3 s^{-1}$ and stacked over 100 random realizations of the Milky Way halo.
\label{angles}}
\end{figure}

Out of the subhaloes entering the histogram of Fig.~\ref{nbclumps}, some are extended. Fig.~\ref{angles} displays the 
distributions of fluxes and angular size of the subhaloes computed over 100 random realizations of the Milky Way for two annihilation cross sections. Here the extension, dubbed $\theta_{90}$, is defined as the angular size on the sky of the regions where 90\% of the annihilation signal arises from. Note that the brightest clumps are always more extended than the average, but they are never larger than about $1.5^{\circ}$. This is important because as it does not conflict with the on-off background subtraction technique discussed in Sec.~\ref{SensitivityMaps}.

\begin{figure}[t]
\centering
\includegraphics[width=.8\columnwidth]{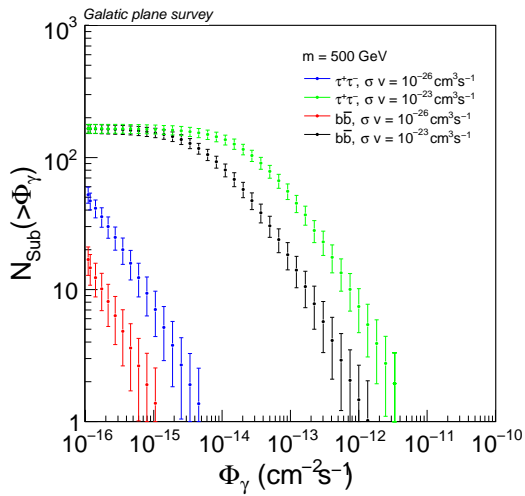}
\caption{Integrated luminosity function for the VL-II subhaloes in the HESS field of view. The considered WIMP mass is 500 GeV; different cases for the annihilation parameters are considered: annihilations into $b\bar{b}$ and $\tau^+\tau^-$ and cross sections of $10^{-23}$ and $10^{-26}\;\rm cm^3 s^{-1}$.
\label{IntegratedLumi}}
\end{figure}

To give an idea of how many of the subhaloes of Fig.~\ref{nbclumps} are visible for a given sensitivity, one can refer to the integrated luminosity function in Fig.~\ref{IntegratedLumi}. It presents the number of subhaloes in the field of view that have a flux larger than a given value versus the value of the flux. The error bars correspond to the rms of the distribution due to the random nature of the halo realization. The results are presented for different particle DM configurations: 500 GeV WIMP with annihilations into $b\bar{b}$ and $\tau^+\tau^-$ and cross sections of $10^{-23}$ and $10^{-26}\;\rm cm^3 s^{-1}$. Should an experiment have an arbitrarily low flux sensitivity, it would lie on the lower-end asymptote and observe all of the 168 objects on average. If the sensitivity map was constant over the whole field of view, to get a constraint one should just set the annihilation cross-section and mass of the DM particle to have 2.3 visible objects on average. In practice, the whole map including its fluctuations of sensitivity has to be used.

Considering that HESS has an angular resolution of 0.01$^{\circ}$, a fraction ($\sim 50\%$) of the clumps that are found to be above the HESS sensitivity are extended. This has to be handled as the sensitivity map is built assuming point like sources. In that case, we make the conservative choice to rescale the flux by lowering it to the value enclosed in the instrument angular resolution. Because the flux sensitivity concerns the integrated flux above some threshold, its value depends on an assumed spectrum. Depending on the considered annihilation channel (hereafter $\chi\chi\rightarrow b\bar{b}$ or $\tau^+ \tau^-$) and on the DM particle mass, the values of the HESS sensitivity in each bin of the map are then properly rescaled.

\begin{figure*}[t]
\centering
\includegraphics[width=.8\columnwidth]{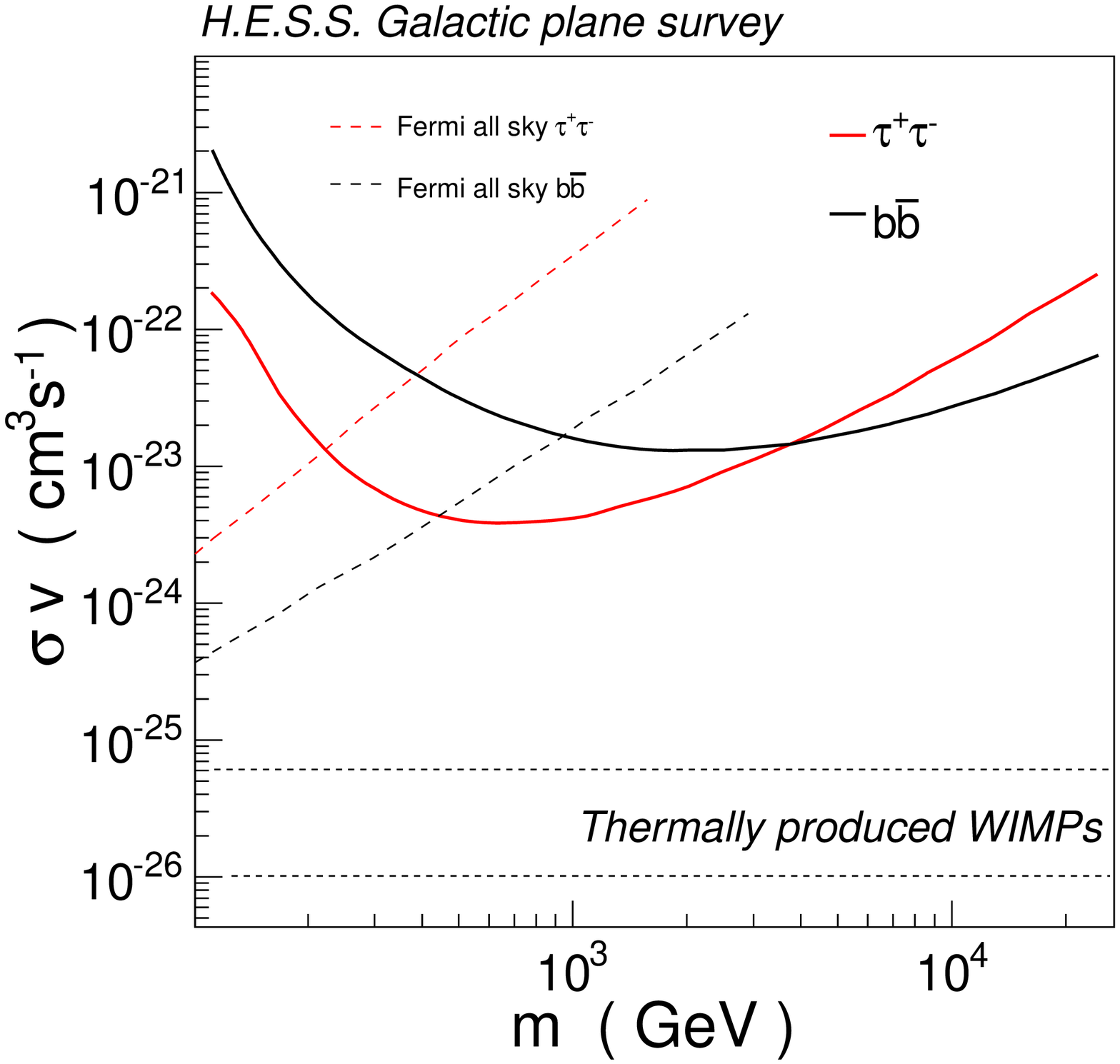}
\includegraphics[width=.8\columnwidth]{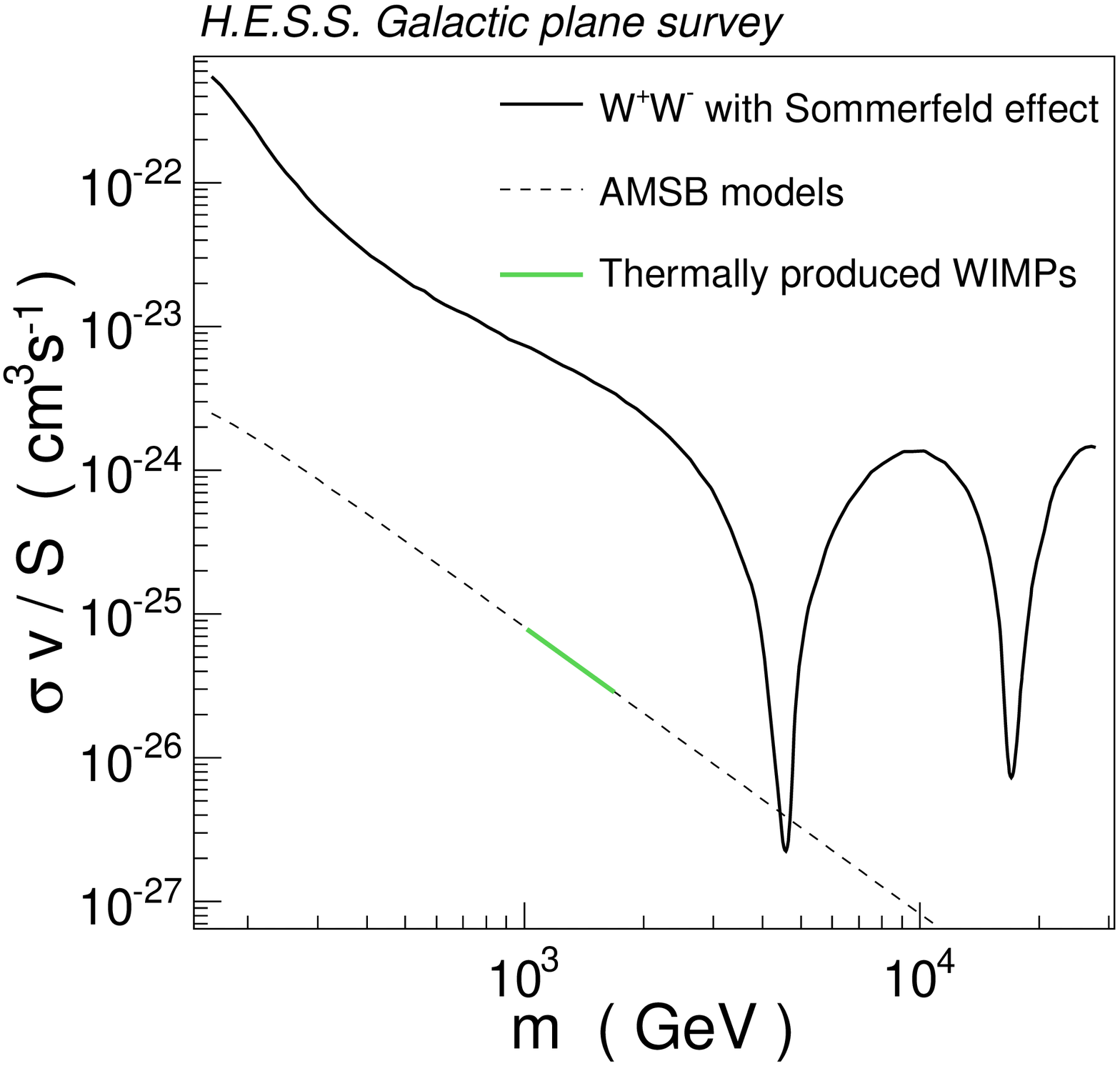}
\caption{Left: Exclusion curves on $\sigma v$  versus the DM particle mass $m$ for  HESS. The limit is calculated at  the 90\% C. L.  for the  DM clumps provided by the VL-II simulation. The DM particle is assumed to annihilate into purely $b\bar{b}$ and $\tau^+\tau^-$ pairs, respectively. The region of natural values of the velocity-weighted annihilation cross section of thermally produced WIMPs is also plotted. Right: Exclusion curves on $\sigma v$  versus the DM particle mass $m$ for HESS including the Sommerfeld enhancement effect.
\label{exclusions}}
\end{figure*}

The left hand side of Fig.~\ref{exclusions} shows the 90\% C.L. exclusion limit on $\sigma v$ as a function of the DM particle mass. Two annihilation spectra are considered: 100\% branching ratio annihilation channel in $b\bar{b}$ and $\tau^+\tau^-$, respectively, in order to somehow encompass all possible annihilation spectra for the DM particle. The limits on the annihilation cross section reach  a few $\rm 10^{-24}\;cm^{3}s^{-1}$ at 1 TeV for the $\tau^+\tau^-$ spectrum. The dashed region corresponds to cosmologically relevant values for the annihilation cross section. The obtained constraints are 2 orders of magnitude above this region.

These results are compared to those obtained with wide-field searches in Fermi data in~\cite{Buckley:2010vg}, which appear to be very complementary as they cover lower energies. 
For the sake of comparison, the best current constraints obtained with Cherenkov telescopes from targeted searches are from the Galactic center with HESS (of the order of $10^{-24}-10^{-23}\;\rm cm^{3}s^{-1}$~\cite{Aharonian:2006wh}) and Sagittarius dwarf galaxy (from $10^{-25}$ to $10^{-23}\;\rm cm^{3}s^{-1}$, depending on the modeling of the source~\cite{Aharonian:2007km}). The constraints obtained here from VL-II subhaloes are based on canonical assumptions and reach  $10^{-24}-10^{-23}\;\rm cm^{3}s^{-1}$; they are thus among the most competitive to date obtained with $\gamma$ rays in this range of mass. 

In the case of an annihilation into gauge boson pairs (here $W^+ W^-$), it could happen that the cross section is significantly increased by means of the so-called Sommerfeld effect~\cite{Hisano:2007zz, ArkaniHamed:2008qn}. The enhancement factor $S$ depends on the mass of the DM particle and on the relative velocity of the colliding particles, it ranges from a few percent up to very large values of $\sim$$10^4$.  For $V_{\max}\ll c$, the enhancement goes approximately as $V_{\max}^{-1}$ before $S$ reaches a plateau due the finite range of the Yukawa interaction. When $m$ is close to a WIMP almost-bound state, the annihilation process becomes resonant with $S\propto V_{\max}^{-2}$. The Sommerfeld effect has been modeled here in the case of an annihilation into $W$ pairs, with an annihilation proceeding towards the exchange of  a 90 GeV boson and a coupling constant of $g=1/30$. In the HESS sensitivity range, the resonances are obtained for $m \sim 4.5\;\rm TeV$ and $m\sim 17.6\;\rm TeV$. In the specific case of annihilations within substructures, the enhancement can become very large, because the colder the subhalo, the larger the enhancement factor~\cite{Lattanzi:2008qa}. In the subhaloes  considered in VL-II,  the maximal velocity $V_{\rm max}$ ranges from $0.5$ to $20\; \rm km\, s^{-1}$. Having $S\propto V_{\max}^{-1}$, the boost factor can indeed be large~\cite{Kuhlen:2009kx}. In our Monte Carlo realizations, a specific Sommerfeld boost is assigned to each subhalo depending on its specific $V_{\rm max}$. The constraints obtained are displayed in the right panel of  Fig.~\ref{exclusions}. Some predictions from supersymmetric models with annihilation into $W$ bosons, extracted from~\cite{Moroi:1999zb} [in the anomaly-mediated supersymmetry breaking scenario (AMSB)], are also shown. These predictions do not include the $S$ factor, so constraints on the unboosted cross section $\sigma v / S$ are shown.  Outside resonances, the limit is less than 2 orders of magnitude above the annihilation cross section expected for thermally produced WIMPS, but --thanks to the resonant Sommerfeld effect-- a small region around 4.5 TeV is excluded.

\section{Prospects for CTA observations programs}
\label{prospects}
\subsection{HESS-like Galactic plane survey}

\begin{figure*}[t]
\centering
\includegraphics[width=.8\columnwidth]{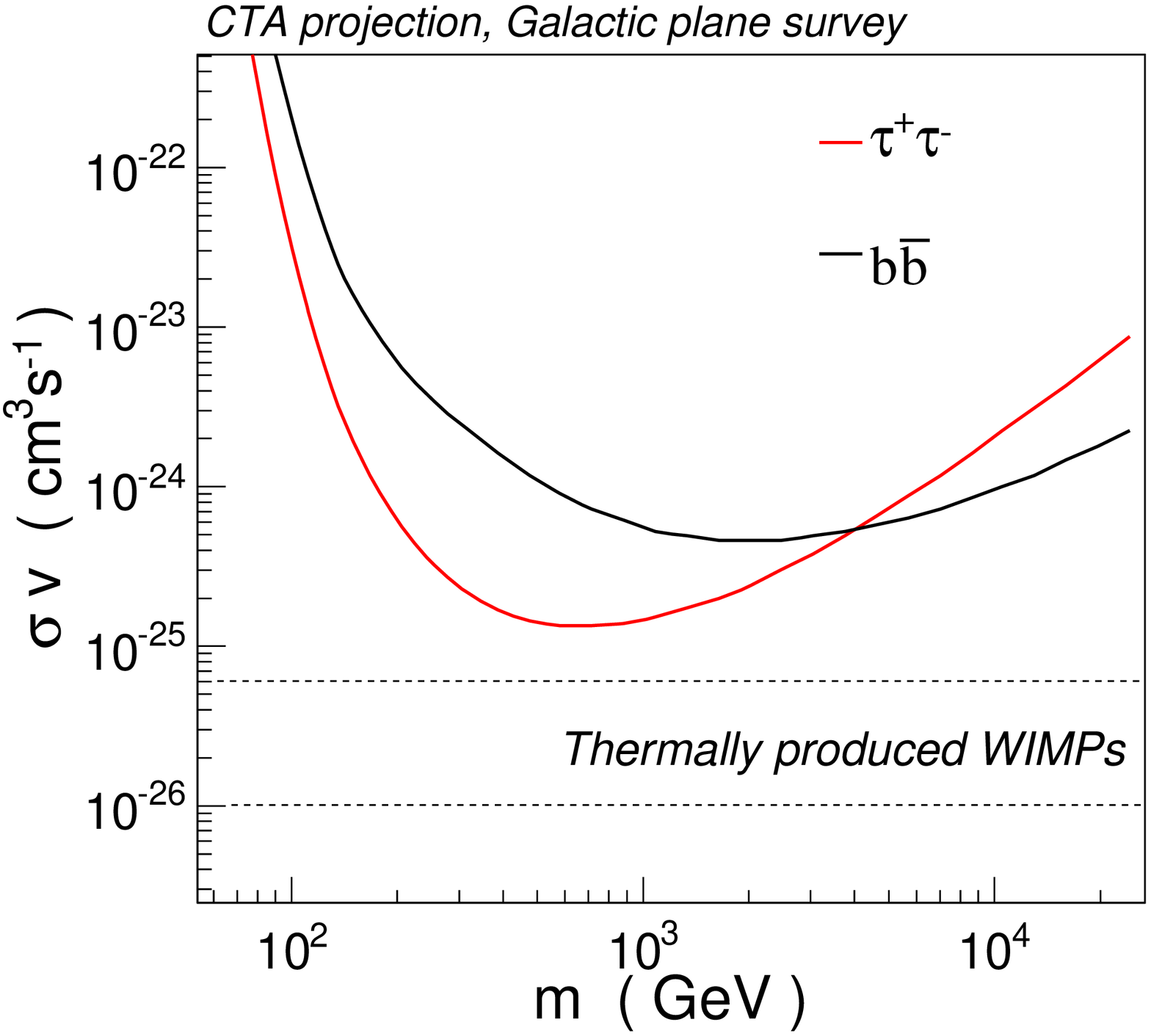}
\includegraphics[width=.8\columnwidth]{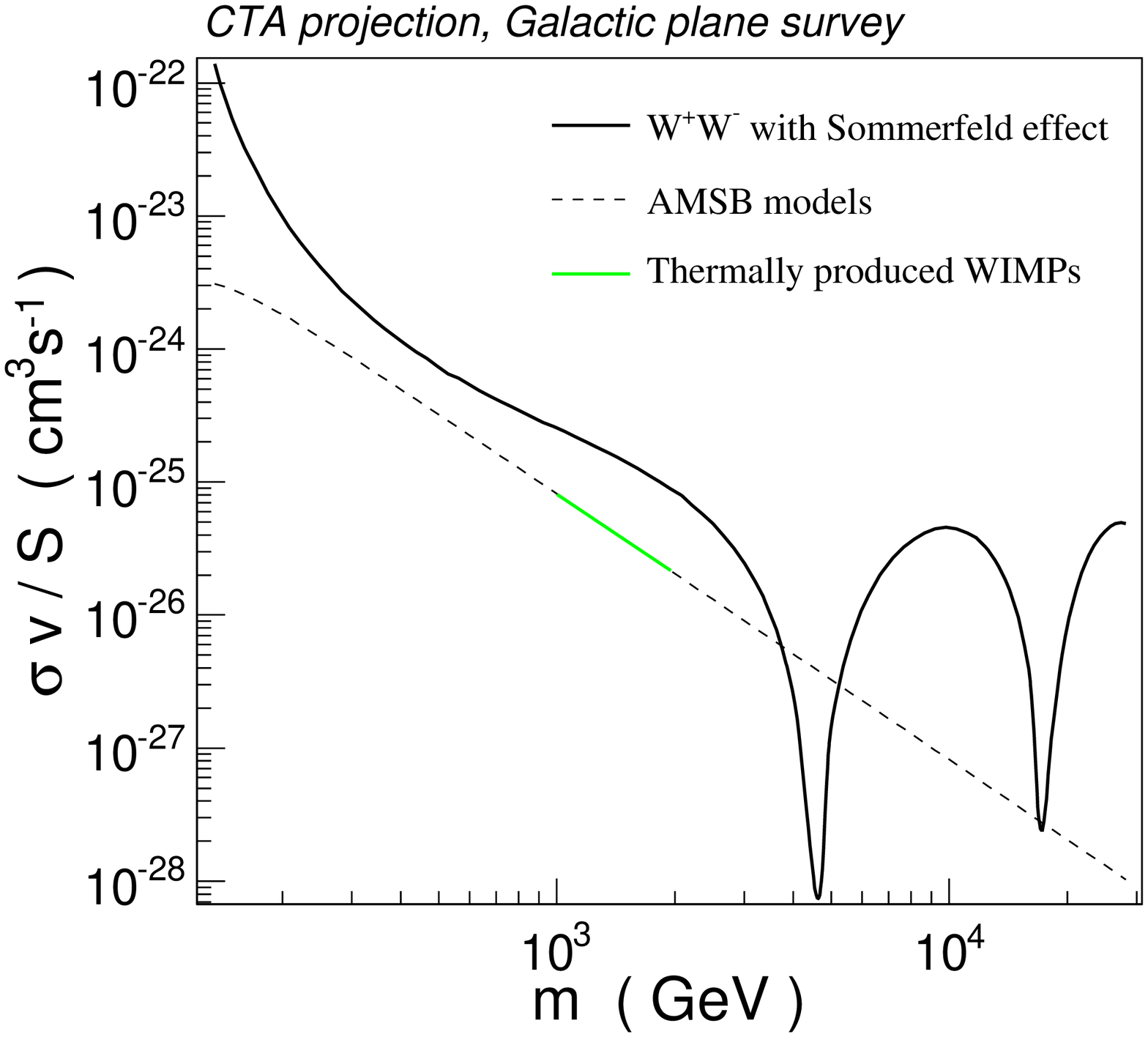}
\caption{Left:  The projected sensitivity curves on $\sigma v$  versus the DM particle mass $m$ for a CTA Galactic plane survey. The limit is calculated at the 90\% C. L. for the  DM clumps provided by the VL-II simulation. The DM particle is assumed to annihilate into purely $b\bar{b}$ and $\tau^+\tau^-$ pairs respectively. The region of natural values of the velocity-weighted annihilation cross section of thermally produced WIMPS is also plotted. Right: Exclusion curves on $\sigma v$  versus the DM particle mass $m$ for a CTA-like array including the Sommerfeld enhancement effect.
\label{exclusions_cta}}
\end{figure*}

The projected map for the CTA is used as in the previous analysis of the HESS Galactic survey to make a prediction for the sensitivity of the future array. As a first step, the same field of view as HESS is used. We consider that  a scan of the Galactic plane will for sure be performed by the CTA, so that this region of the sky is somehow the minimal guaranteed field of view.
An exposure of 10 h in each pixels corresponds to a total observation time for building up the survey of 400 h.
Concerning the extension of the sources, a slightly more optimistic method is used. Instead of rescaling the flux to the signal enclosed in the angular resolution, the whole signal is considered. Nevertheless, any clump that would be too much extended to allow for a proper background subtraction is excluded from the sample. Subhaloes with $\theta_{90}>1.5^{\circ}$ are thus not considered.
The results for the projection to the CTA are presented in Fig.~\ref{exclusions_cta}. The exclusion limits are lower by a factor of $\sim$10  than those obtained with HESS. In the conventional case ($b\bar{b}$ and $\tau^+\tau^-$, without Sommerfeld enhancement), they are reaching $\sigma v$ values of a few 10$^{-25}$cm$^{3}$s$^{-1}$. In the case of Sommerfeld enhanced  annihilations, some regions of the parameter space for the model could be excluded,  since a large array of telescopes would have enough sensitivity to detect WIMPs in the mass range from $\sim$3 to 6 TeV and close to the second resonance. 

We conclude from Fig.~\ref{exclusions_cta} that using this field of view, the CTA will not be able to reach signals from the most natural WIMPs. One order of magnitude is gained with respect to HESS, but a factor of 2--10  is still necessary to  reach the natural DM annihilation cross sections. An homogeneous increase of the exposure time will only improve the exclusion limits as the square root of exposure time in the background-limited regime, so one has to enlarge the field of view instead of using longer exposure. In addition, the flux sensitivity  along the Galactic plane will be limited by the population of  newly detected sources at  a flux level of 10$^{-12}$ cm$^{-2}$s$^{-1}$. The Galactic plane might  also not be the best place to look for subhaloes since they could have been tidally affected by the disk. For those reasons, an observing strategy focusing  on fields with absolute Galactic latitude of at least 0.5$^{\circ}$ should be preferred for DM subhalo searches, as it clearly appears in the lower panel of Fig.~\ref{maps}. This is precisely the point developed in the next subsection.

\subsection{CTA survey of one-quarter of the sky}

\begin{figure}[t]
\centering
\includegraphics[width=.8\columnwidth]{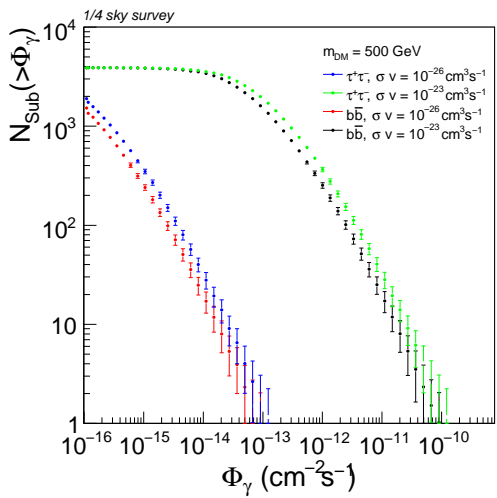}
\caption{Integrated luminosity function for the VL-II subhaloes in the field of view of one quarter of the sky. The considered WIMP mass is 500 GeV; different cases for the annihilation parameters are considered: annihilations into $b\bar{b}$ and $\tau^+\tau^-$ and cross sections of $10^{-23}$ and $10^{-26}\;\rm cm^3 s^{-1}$
\label{IntegratedLumi_CTA}}
\end{figure}

To go further, one can note that larger scans of the sky will most likely be  conducted by the CTA.  In particular, a more extended survey of the order of a quarter-sky size is foreseen. In this section, the CTA sensitivity  to DM annihilations is computed in the context of such an ambitious program. A large survey increases the probability to find bright subhaloes in the field of view, which thus translates into better constraints. 
 Such a survey should not include the Galactic plane where numerous sources are expected to shine and therefore decrease the sensitivity  to DM clumps. On the other hand, the central region of the Milky Way is attractive since the VL-II subhalo distribution is peaked towards the center. For this study, the survey region is chosen to be from -90$^{\circ}$  to +90$^{\circ}$ in Galactic longitude and from -45$^{\circ}$  to +45$^{\circ}$ in Galactic latitude, excluding the Galactic plane between $\pm$1.5$^{\circ}$.  Inside this region, the distribution of the number of subhaloes from the simulation is similar to the one of Fig.~\ref{nbclumps}, but with an average value of 3907, and a rms of 324. 
The fact that this extended survey leads to better constraints is illustrated in Fig.~\ref{IntegratedLumi_CTA}, where the same results as for Fig.~\ref{IntegratedLumi} are presented, but for the considered quarter-sky survey instead of the Galactic plane survey. Here again, the quoted error bars correspond to the RMS of the distribution over possible realizations.

\begin{figure*}[t]
\centering
\includegraphics[width=.8\columnwidth]{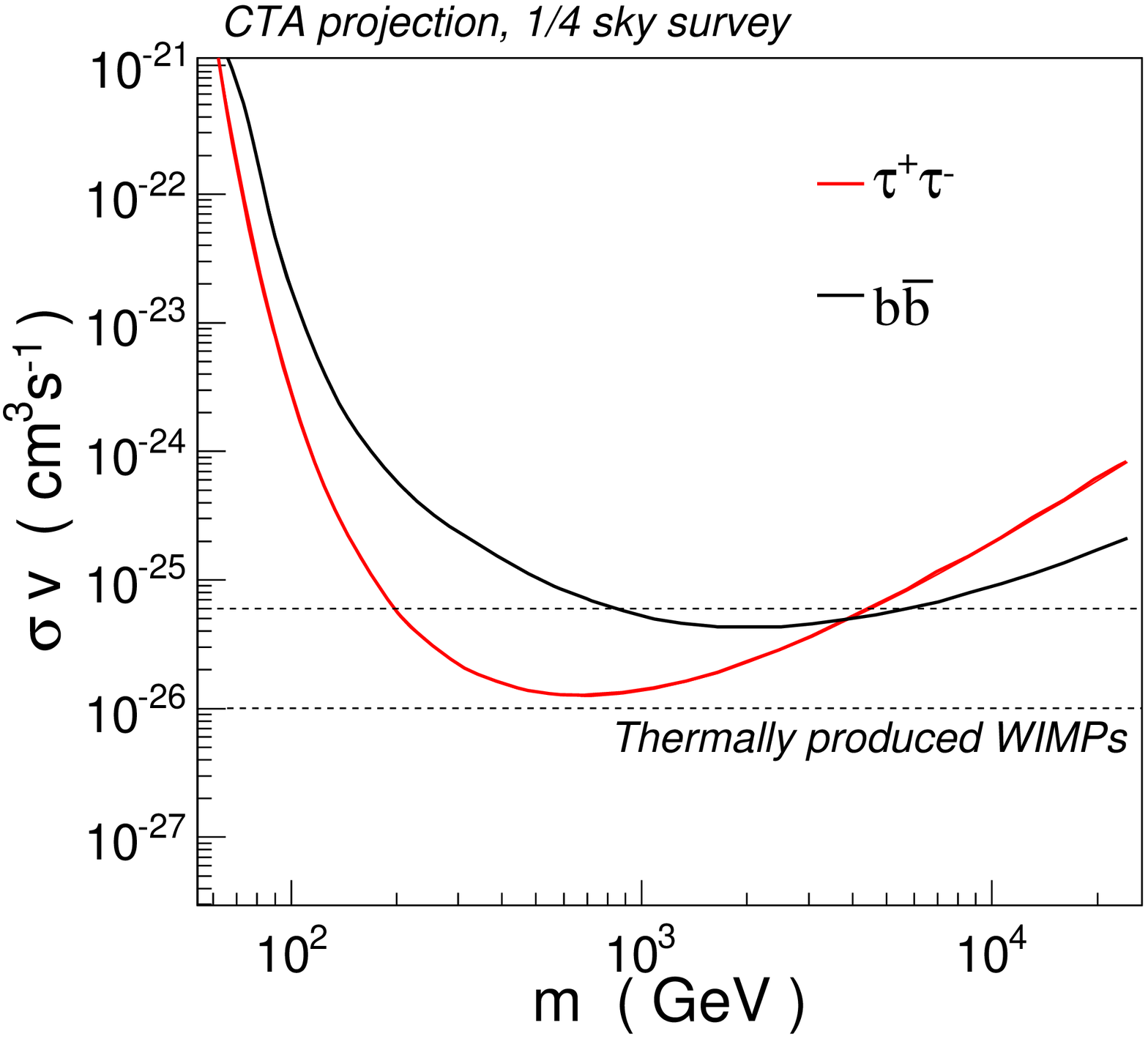}
\includegraphics[width=.8\columnwidth]{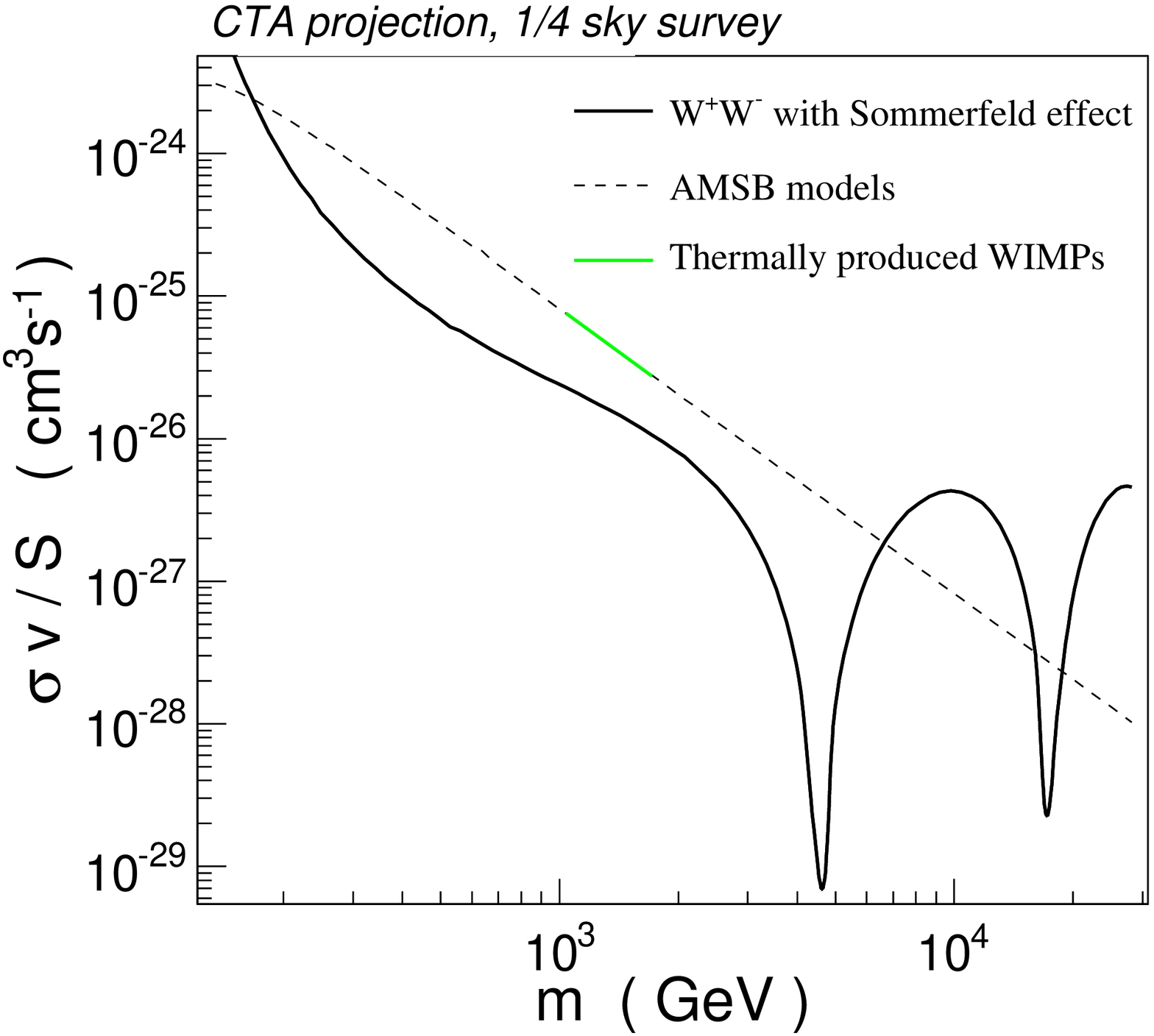}
\caption{Left: Exclusion curves on $\sigma v$  versus the DM particle mass $m$ for a  CTA survey of a one fourth of the sky. The limit is calculated at 90\% C. L. for the DM clumps provided by the VL-II simulation. The DM particle is assumed to annihilate into purely $b\bar{b}$ and $\tau^+\tau^-$ pairs respectively. The region of natural values of the velocity-weighted annihilation cross section  of thermally produced WIMPs is also plotted. Right: Exclusion curves on $\sigma v$  versus the DM particle mass $m$ for the CTA including the Sommerfeld enhancement effect.
\label{exclusionsCTAquart}}
\end{figure*}

 The sensitivity is taken to be constant on the entire field of view. Its value is calculated from the previous CTA sensitivity map averaged for Galactic latitudes above 1.5$^{\circ}$ and corrected for a shorter exposure. A 5h exposure time in each pixel leads to a flux sensitivity of the order of 5$\times$ 10$^{-13}$cm$^{-2}$s$^{-1}$ for  a WIMP mass of 500 GeV.
For this last study as well, the whole signal from extended subhaloes is considered, and objects larger than $1.5^\circ$ are rejected.
As in previous sections, the value of the sensitivity is renormalized for each DM particle mass. To a very good approximation, the decrease of the sensitivity due to the new population of extragalactic sources such as, {\it e.g.} active galactic nuclei, is negligible. The quoted value of the sensitivity is reached by pointing the whole array in the same direction, which is quite time consuming.  Assuming a duty cycle of 1000 h of observation per year, this quarter-of-the-sky survey can be completed within about 6 years of operation. An implicit assumption is that all unidentified sources do not present the required features of DM source candidates. As for the Galactic survey, the observation of extension, variability, and energy spectrum will be used to rule out the unidentified sources as DM clump candidates.
Notice however that should a couple of plausible DM sources be detected, additional dedicated observation time would be devoted for deeper studies.
 
Figure~\ref{exclusionsCTAquart} shows the 90\% C.L. exclusion limit on $\sigma v$ as a function of the DM particle mass. Annihilation cross sections of a few 10$^{-26}$ cm$^{3}$s$^{-1}$ are reached in the 200 GeV - 3 TeV mass range in the case of annihilation into $\tau$ pairs. In the scenario of an annihilation into $W$ bosons pairs with Sommerfeld enhancement, all WIMPs from the AMSB with masses from 200 GeV to 6 TeV are within the reach of the CTA.

\section{Conclusions}
\label{Conclusions}

We used for the first time a wide field-survey from Cherenkov telescopes to constrain conventional DM substructure scenarios. Unlike the case when DM annihilations are searched for towards selected sources, the constraints from this blind search do not rely crucially on the modeling of the DM distribution in the source. The constraints obtained out of the HESS Galactic plane survey are still 2 orders of magnitude higher than the thermal WIMP region.  Thus most natural models for WIMPs as dark matter are out of the reach of current generation ground-based Cherenkov telescopes with wide-field surveys and realistic observing time. However, the limits  reached in the $\sigma v - m$ plane are very competitive compared to other 
strategies such as targeted searches. By using the same Galactic plane field of view, we show that the discovery of particle DM in the form of WIMPs is unlikely to be accessible for the next generation of Cherenkov telescopes such as the CTA. However, by considering an ambitious but realistic quarter-of-the-sky survey with the CTA, it is shown that the thermal WIMP promised land can be hit. Note that such a survey will likely be conducted by CTA independently of particle DM considerations. This search for DM subhaloes will therefore not be in conflict with other physics programs.

\begin{acknowledgments}
JD acknowledges support from the Swiss National Science Foundation (SNSF).
\end{acknowledgments}

\bibliography{bmdg}


\end{document}